\documentclass[9pt]{extarticle}
\usepackage{spconf,amsmath,graphicx}
\usepackage[colorinlistoftodos]{todonotes}
\usepackage{caption}
\usepackage{subcaption}
\usepackage{bm}
\usepackage{hyperref}
\usepackage{xcolor}
\usepackage{color}
\usepackage{bm}
\usepackage{url}
\usepackage{subcaption}
\usepackage{booktabs}
\usepackage{multirow}
\usepackage{textcomp}
\usepackage{verbatim}
\usepackage{amssymb}
\hypersetup{
    colorlinks=true,
    linkcolor=blue,
    filecolor=magenta,      
    urlcolor=blue,
}

\newcommand{\be}[1]{\textbf{#1}}
\newcommand{\se}[1]{\underline{#1}}

\title{Guided Speech Enhancement Network}
\name{Yang Yang, Shao-Fu Shih, Hakan Erdogan, Jamie Menjay Lin, Chehung Lee, Yunpeng Li, George Sung, Matthias Grundmann}

\address{Google LLC, U.S.A.}

\begin{document}
%
\maketitle
\begin{abstract}
High quality speech capture has been widely studied for both voice communication and human computer interface reasons. To improve the capture performance, we can often find multi-microphone speech enhancement techniques deployed on various devices. Multi-microphone speech enhancement problem is often decomposed into two decoupled steps: a beamformer that provides spatial filtering and a single-channel speech enhancement model that cleans up the beamformer output. In this work, we propose a speech enhancement solution that takes both the raw microphone and beamformer outputs as the input for an ML model. We devise a simple yet effective training scheme that allows the model to learn from the cues of the beamformer by contrasting the two inputs and greatly boost its capability in spatial rejection, while conducting the general tasks of denoising and dereverberation. The proposed solution takes advantage of classical spatial filtering algorithms instead of competing with them. By design, the beamformer module then could be selected separately and does not require a large amount of data to be optimized for a given form factor, and the network model can be considered as a standalone module which is highly transferable independently from the microphone array. We name the ML module in our solution as GSENet, short for Guided Speech Enhancement Network. We demonstrate its effectiveness on real world data collected on multi-microphone devices in terms of the suppression of noise and interfering speech.

\end{abstract}
\vspace{-0.2cm}
\section{Introduction}
\label{sec:intro}

\begin{figure}[!t]
 \centering
 \begin{subfigure}[b]{0.45\textwidth}
     \centering
     \includegraphics[width=\textwidth]{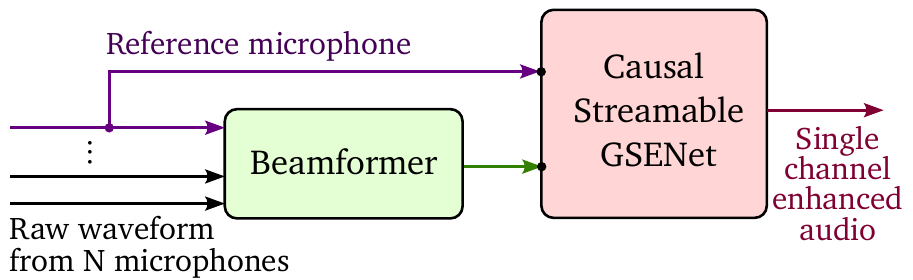}
     \caption{System diagram. GSENet takes both beamformer output as well as waveform from raw reference microphone as input.}
     \label{fig:diagram}
 \end{subfigure}
 \hfill
 \vspace{0.01em}
 \begin{subfigure}[b]{0.5\textwidth}
     \centering
     \includegraphics[width=\textwidth]{plots/polar_teaser_camera_ready.pdf}
     \caption{BSS-SDR (dB) with interfering speech from 8 directions\footnotemark, each at 0\,dB SNR on a device with 3 microphones. SSENet, short for Single-microphone Speech Enhancement Network, is a baseline speech enhancement network with only the beamformer output as input. Target speaker is at 0\,\textdegree. A fixed LTI beamformer is used. See Fig.~\ref{fig:listening_room} for the lab recording setup, Table~\ref{table:bss_sdr} for more complete results, and Section~\ref{sec:exp} for details about models and experiment setup.}
     \label{fig:result_highlight}
 \end{subfigure}
\caption{System diagram and result highlight.}
\vspace{-0.5cm}
\label{fig:teaser}
\end{figure}
\footnotetext{The values in between the 8 discrete angles are interpolated by periodic cubic spline for better visualization. Note that this is not the typical directivity pattern plot, but the pattern for spatial rejection. The stronger the spatial rejection, the higher the BSS-SDR values.}

Voice is one of the primary methods of communication between individuals. As technology advances, speech as a computer interface has also become essential to human-machine interaction. For reliable communication, a robust audio front-end is needed to enhance speech and combat various sources of degradation such as echo, background noise, interfering speech, and room reverberations.

The importance of speech enhancement and the renaissance of deep learning have motivated a plethora of research on learning-based speech enhancement solutions, from speech denoising and dereverberation \cite{ tan_gcrn_trans_aslp_2020, hu_dccrn_interspeech_2020, luo_convtasnet_trans_aslp_2019, gfeller_denoise_music_ismir_2020, li_sdd_interspeech2021} to bandwidth restoration and extension \cite{li_seanet_bwe_icassp_2021, lee_nu_wave_interspeech_2021} to unified approaches \cite{liu_voicefixer_2021}. Most of these solutions, however, focus on the case with single-channel input. 

To improve the microphone capture quality, multi-micro\-phone solutions are commonly used to increase the system SNR and spatial resolution \cite{gannot_trans_aslp_2017}. Hence, most smartphones, tablets, laptops, and smart speakers are now equipped with multiple microphones. The availability of multi-channel audio input provides spatial diversity and promises better separation of speech and interference \cite{gannot_trans_aslp_2017}. Multi-mic speech enhancement is often decomposed into two decoupled stages: a beamformer that provides spatial filtering, the design of which is specific to the microphone geometry, and a single-channel speech enhancement model that cleans up the beamformer output. Recent research has shown the possibility of an end-to-end multi-channel model \cite{luo2019fasnet,luo2020end,lee2019end} especially in the context of speech separation. In our task, however, end-to-end neural approaches tend to suffer from the requirements for large device-specific data collection effort and device-specific re-training and optimization, which is challenging when scaling. In addition, simulation-based training of multichannel models does not generalize well to real-world conditions due to acoustic mismatches. The question we ask is: Can we devise a speech enhancement solution that takes advantage of the multi-microphone input  while allowing the neural network component to be agnostic to microphone array configurations and easily transferable between devices and acoustic environments?


In this work, we consider the scenario where there is a single target speaker at a given spatial location relative to a multi-microphone device, and the task is to reject interfering speech and/or background noise that come from various directions. Instead of decomposing the problem into decoupled spatial filtering and single-channel speech enhancement steps, or opting for end-to-end learning-based solutions that require device-specific training data, we propose a solution comprised of a classical beamformer and a dual-input speech enhancement neural network model. As illustrated in Fig.~\ref{fig:teaser}(a), the model takes both the beamformer output and a pre-selected reference raw microphone based on which the beamformer is derived.

The core to our solution is a device-agnostic training scheme that allows the model to learn from the cues of the beamformer by contrasting the two input sources and enhances its spatial rejection. As contributions, the proposed solution provides (1) a novel end-to-end integration of a classical spatial filtering algorithm and a learning-based speech enhancement model; (2) joint spatial rejection and denoising capability within a single model, capable of real-time streaming inference; and (3) a significant boost in spatial rejection, shown empirically, at challenging low-SNR speech interference scenarios with only three microphones.


As related works, \cite{li_embed_and_beamform_icassp_2022, zhang_adl_mvdr_icassp2021, heymann_nn_based_mask_for_acoustic_beamforming_icassp_2016} propose the use of neural network models to derive linear beamformer coefficients, whereas in our approach the model
learns from the contrast between the input and the output of an existing beamformer to obtain better spatial rejection with non-linear processing. \cite{chen_location_guided_slt2018} addresses the problem of speech separation when an external information provides an azimuth angle and uses an ``angle feature'' derived from the steering vector corresponding to that angle to guide the separation, whereas we do not require an angle or steering vector and instead we use a fixed beamformer estimated from data. The problem of speech separation for ASR is addressed in \cite{omealley_asr_frontend_interspeech2022} and \cite{caroselli_cleanformer_2022}, where the former makes use of a speaker embedding for conditioning, and the latter relies on an adaptive noise suppression algorithm as context. In contrast, our solution is a general speech enhancement algorithm aimed to increase audibility instead of recognition type tasks which are optimized against word error rate.

\vspace{-0.4cm}
\section{Method}
\vspace{-0.2cm}
\label{sec:method}

One of the challenges for end-to-end ML-based processing is the heavy dependency on quality data collection. These approaches are highly prone to issues with model training stability and multi-device training-to-deployment lead time. To address the performance consistency and scalability concerns, we propose a solution that contains a device-specific beamformer module and a machine learning module as illustrated in Fig.~\ref{fig:teaser}(a). The pipeline is designed to take $N$ raw microphone channels as input to the beamformer. The network component, termed GSENet (short for ``guided speech enhancement network''), is a causal streamable network that takes both reference microphone audio and beamformer output as inputs.
The goal of having both input sources is to enable the network to observe the effect of beamforming and then perform inference by comparing the beamformer output and the raw microphone to further reduce the directional interference. With proper training, the network should be able to contrast the two inputs, pick up the spatial contrast depicted between the two then enhance the directionality based on spectral rejection. 


\subsection{Data-Pipeline and Training}

To train a model that is agnostic to different microphone array configurations and beamformer implementations, we choose to simulate the effect of beamforming with synthesized data by controlling the gains of different signal components. 

Let us denote $\bm{s}, \bm{n},$ and $\bm{i}$ as the waveform of the target speech, noise, and interfering speech, respectively. A room simulator is used to sample the room impulse responses (RIR) between three signal sources and two microphones, with randomly sampled room layout and source and receiver locations for each training step. We denote the RIRs from source $k$ to receiver $j$ as $\bm{r}_{(k, j)}$. The two input channels to the model are synthesized as
\begin{align}
    \bm{y}_0 = &  \bm{s} * \bm{r}_{(0,0)} + g_n \cdot\bm{n} * \bm{r}_{(1,0)}  + p_i \cdot g_i \cdot \bm{i} * \bm{r}_{(2,0)}, \text{ and} \notag\\
    \bm{y}_1 = & \bm{s} * \bm{r}_{(0,1)} +  \alpha\cdot g_n \cdot\bm{n} * \bm{r}_{(1,1)} + \beta \cdot p_i \cdot g_i \cdot \bm{i} * \bm{r}_{(2,1)}, \notag
\end{align}
where $*$ denotes temporal convolution. The power of $\bm{s}, \bm{n}, \bm{i}$, and $\bm{r}_{(k,j)}$ are normalized. $g_n$ and $g_i$ are random scalar gains common to both inputs that define the relative strength of the noise and inference compared with the target speech\footnote{There is also a global gain sampled and applied to all signal components. We omit it here for brevity.}. $p_i$ is a Bernoulli random variable modeling the existence of an interfering speech signal $\bm{i}$.

We treat $\bm{y}_0$ as the simulated beamformer output and $\bm{y}_1$ as the simulated reference microphone input, where the effect of beamforming is captured by two  random scalars $\alpha$ and $\beta$\footnote{Partly, the effect of the beamformer is also captured through the difference of $\bm{r}_{(j, 0)}$ and $\bm{r}_{(j, 1)}$, which introduces phase and amplitude differences in $\bm{y}_0$ and $\bm{y}_1$.} with values larger than 1. The assumption here is that both the noise $\bm{n}$ and the interfering speech $\bm{i}$ are attenuated by a beamformer. Thus, the raw microphone input, simulated by $\bm{y}_1$, sees a stronger signal strength of the two than the simulated beamformer output $\bm{y}_0$. During inference we substitute the two input channels $\bm{y}_0$ and $\bm{y}_1$ with the output of a beamformer and the corresponding reference microphone input. 

A few remarks are in order: (a) This formulation assumes that the beamformer does not attenuate or amplify the target signal, which is a common design goal for widely adopted beamforming algorithms like MVDR \cite{erdogan_mvdr_masking_interspeech16}. (b) The effect of a real beamformer is far more complex than captured in this formulation -- it introduces specific patterns of amplitude and phase change as a function of the angle of arrival and signal frequency. Empirically though, we observe that models trained with this data pipeline work well with real beamformers. (c) Since we do not expect a beamformer to introduce sample delays, in the synthetic room simulator we confine the two receivers in close proximity with each other so as not to introduce large sample offsets. 

The training target $\bm{y}_{t}$ is obtained by convolving the target speech $\bm{s}$ with $\bm{r}_{(0,0)}^{\text{anechoic}}$, the anechoic\footnote{The anechoic version of the RIR captures only the strongest path.} version of $\bm{r}_{(0,0)}$. In other words, the model is trained to obtain a denoised and dereverberated version of the target speech.
\vspace{-0.2cm}
\subsection{Model Architecture}

\begin{figure}[!t]
\centering
\includegraphics[width=0.49\textwidth]{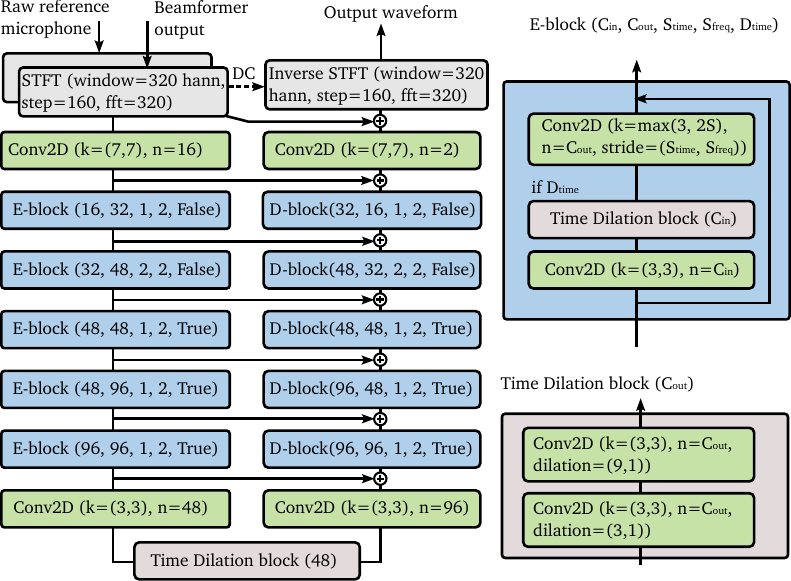}
\caption{GSENet network architecture. All convolutions are causal. Decoder block (D-block) is symmetric to encoder block (E-block). Input and output are 16kHz. We name the single-channel variant of this architecture that takes only the beamformer output as input to be SSENet, short for single-mic speech enhancement network. }
\vspace{-0.3cm}
\label{fig:network}
\end{figure}

Our work utilizes an U-Net design which has a similar overall architecture as the one introduced in  \cite{gfeller_denoise_music_ismir_2020}, except that all convolutions are changed to be causal for real-time streaming inference. Details of the model architecture are shown in Fig.~\ref{fig:network}. 
Throughout the model leaky-ReLU(0.3) is used as activation. The complex STFT outputs are packed as real and imaginary channels. 
A single-scale STFT reconstruction loss \cite{engel_ddsp_iclr_2020} with window size of 1024 and step size of 256 is used during training.

\subsection{Real-time Inference}

For online inference, we leverage the streaming-aware network modules introduced in \cite{raybakov_streamable_interspeech_2020} and extended in
\cite{li_seanet_bwe_icassp_2021}. The streamable modules \cite{kws_streamable} add ring buffers to regular convolutions to keep track of relevant input history so that the network inference can be done incrementally in time without losing history context. As detailed in Fig.~\ref{fig:network}, the innermost layer of the U-Net downsamples STFT frame in time by a factor of 2, which limits the minimum time granularity of streaming inference to once every two STFT frames. Given the STFT window size of 20\,ms and step size of 10\,ms, the two-frame-per-inference limit translates into 30\,ms latency incurred by the model. Profiling the model on a single CPU core of a Pixel 6 mobile phone indicates a processing time of 1.81\,ms for each 20\,ms chunk of audio, giving a total latency of 31.81\,ms. 
Note that typical end-to-end VoIP latency is about 200\,ms so our proposed solution has minimum impact on the existing pipelines. 




\vspace{-0.2cm}
\section{Experiments}
\label{sec:exp}

\begin{figure}[!t]
  \begin{minipage}[c]{0.2\textwidth}
    \includegraphics[width=\textwidth]{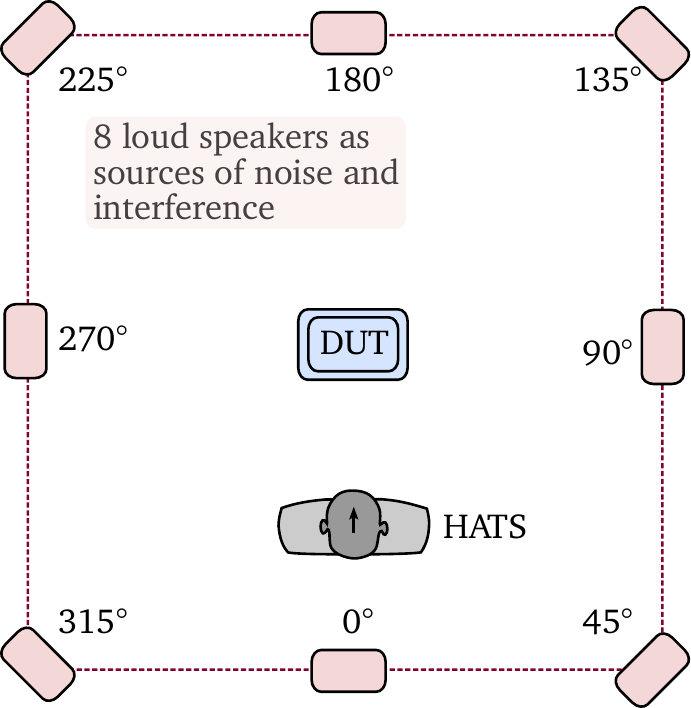}
  \end{minipage}\hfill
  \begin{minipage}[l]{0.25\textwidth}
    \caption{
      Overhead view of the recording setup for evaluation data collection (not drawn to scale). The device under test (DUT) is a 3-microphone device. Speech and noise are played back from the 8 surrounding loudspeakers as interference, while a loudspeaker on the HATS (head and torso simulator) simulates the target speaker. 
    } \label{fig:listening_room}
  \end{minipage}
  \vspace{-0.4cm}
\end{figure}

To validate the performance of our proposed solution, we conduct experiments on GSENet as well as two variants of the corresponding single channel network SSENet(A), and SSENet(B), with training parameters listed in Table~\ref{table:models}. In this setup, SSENet(A) is a typical denoiser trained to separate speech from non-speech; SSENet(B) is trained to additionally reject competing speech with a weaker strength (the mean of $g_i$ is $-9$dB). 
\begin{table}[h]
\setlength{\tabcolsep}{7pt}
\small
\caption{Training parameter setup. $\text{dB}(\cdot)\triangleq 20\log_{10}(\cdot)$.}\label{table:models}\vspace{-0.2cm}
\centering
\begin{tabular}{ @{}l l @{}} 
\toprule
Common  & $\text{dB}(g_n)\sim \mathcal{N}(-5, 10)$ \\
 
\midrule
SSENet (A)  & $p_i=0$\\ 
\hline
SSENet (B)  & $p_i\sim \text{Bernoulli(0.4)}$\\ 
&$\text{dB}(g_i)\sim \mathcal{N}(-9, 3)$ \\
\hline
GSENet      & $p_i\sim \text{Bernoulli(0.4)}$\\
  &$\text{dB}(g_i)\sim \mathcal{N}(-3, 3)$\\
  &$\text{dB}(\alpha)\sim \max(\mathcal{N}(0, 3), -4)$\\
  &$\text{dB}(\beta)\sim \max(\mathcal{N}(4, 6), 4)$\\
\bottomrule
\end{tabular}
\vspace{-0.2cm}
\end{table}

Both target speech $\bm{s}$ and interference speech $\bm{i}$ are sampled from a dataset derived from LibriVox \cite{librivox} together with an internal speech dataset. Background noise $\bm{n}$ is sampled from a dataset derived from Freesound \cite{freesound}. The $3\times2$ Room RIRs $\bm{r}_{(k, j)}$ are generated with randomly sampled rooms using image method \cite{allen_image_method_1979}.

\begin{table*}[t]
\setlength{\tabcolsep}{3.6pt}
\caption{
 BSS-SDR (dB) \cite{raffel_bss_sdr_2014} of the raw and enhanced speech waveform with interference coming from different angles. Left: speech as interference. Right: noise as interference. Top: interference at 0dB SNR. Bottom: interference at 6dB SNR.
}
\label{table:bss_sdr}\vspace{-0.2cm}
\centering

\begin{tabular}{ @{}l c c c c c c c c c c c c c c c c c c c @{}} 
\toprule
& \multicolumn{9}{c}{Speech (VCTK) as interference} & \multicolumn{9}{c}{Noise (DEMAND) as interference}\\
 \cmidrule(lr){2-10} \cmidrule(lr){11-19}
Interference angle   & 0\textdegree & 45\textdegree & 90\textdegree & 135\textdegree & 180\textdegree & 225\textdegree & 270\textdegree & 315\textdegree & avg.  & 0\textdegree & 45\textdegree & 90\textdegree & 135\textdegree & 180\textdegree & 225\textdegree & 270\textdegree & 315\textdegree & avg.\\

\midrule
Reference Mic.  & -0.3 & 0.3 & 0.6 & -0.0 & -0.5 & -0.6 & -0.9 & -0.2 & -0.2  & 0.3 & 0.4 & 0.7 & 0.3 & -0.5 & -0.4 & -0.4 & -0.3 & 0.0\\ 
Beamformer(BFer)  & 1.4 & 2.1 & 2.3 & 1.8 & \se{0.4} & 2.1 & 2.2 & 2.3 & 1.8 & 5.6 & 8.1 & 6.4 & 5.1 & 4.1 & 6.2 & 7.0 & 6.4 & 6.1 \\ 
BFer + SSENet (A) & 1.7 & 2.2 & 2.3 & 1.9 & \se{0.4} & 2.2 & 2.3 & 2.5 & 1.9  & 11.3 & 12.7 & 12.0 & 10.7 & \se{10.1} & 11.8 & 12.1 & 11.7 & 11.5\\
BFer + SSENet (B) & \se{2.8} & \se{3.1} & \se{3.1} & \se{2.7} & -1.0    & \se{3.0} & \se{3.1} & \se{3.5} & \se{2.5}   & \be{11.8} & \se{13.1} & \se{12.5} & \se{11.3} & \be{10.9} & \se{12.3} & \se{12.6} & \se{12.2} & \se{12.1} \\ 
BFer + GSENet    & \be{6.5} & \be{9.2} & \be{9.7} & \be{8.5} & \be{2.0} & \be{9.1} & \be{10.7} & \be{10.1} & \be{8.2}  & \se{11.4} & \be{13.3} & \be{12.6} & \be{11.5} & 10.0 & \be{12.7} & \be{13.1} & \be{12.7} & \be{12.2}\\ 

\midrule

Reference Mic.    & 5.5 & 6.1 & 6.4 & 5.7 & 5.3 & 5.2 & 5.0 & 5.6 & 5.6 & 6.1 & 6.2 & 6.4 & 6.1 & 5.3 & 5.4 & 5.4 & 5.5 & 5.8\\ 
Beamformer(BFer)  & 7.2 & 7.7 & 7.9 & 7.5 & 6.2 & 7.8 & 7.9 & 7.9 & 7.5 & 10.7 & 12.5 & 11.4 & 10.3 & 9.4 & 11.2 & 11.8 & 11.3 & 11.1\\
BFer + SSENet (A)  & 7.3 & 7.8 & 7.9 & 7.5 & 6.1 & 7.8 & 7.8 & 8.1 & 7.5 & \se{14.5} & 15.3 & 15.0 & 14.1 & \se{13.6} & 14.9 & 15.1 & 14.8 & \se{14.7}\\ 
BFer + SSENet (B) & \se{9.3} & \se{9.7} & \se{9.8} & \se{9.4} & \be{7.5} & \se{9.8} & \se{9.8} & \se{10.1} & \se{9.4} & \be{14.8} & \se{15.6}.    & \se{15.3} & \se{14.5} & \be{14.1} & \se{15.1} & \se{15.4} & \se{15.0} & \be{15.0}\\
BFer + GSENet   & \be{10.2} & \be{12.5} & \be{12.8} & \be{12.1} & \se{7.0} & \be{12.5} & \be{13.6} & \be{13.3} & \be{11.8}& \se{14.5} & \be{15.8} & \be{15.4} & \be{14.7} & 13.3 & \be{15.4} & \be{15.7} & \be{15.5} & \be{15.0}\\

\bottomrule
\end{tabular}

\end{table*}

To test the performance of our solution in a real world scenario, we collect recordings in listening rooms using a 3-microphone device and build an LTI beamformer specific to the microphone geometry, as detailed next.

\vspace{-0.3cm}
\subsection{Multi-microphone Evaluation Dataset}
\vspace{-0.2cm}


Evaluation datasets are collected in a ETSI certified listening room using the setup illustrated in Fig.~\ref{fig:listening_room}. The DUT in the middle is a 3-microphone device from which the recording is taken. In this setup, there are a total of 9 speakers, including one as the Head and Torso Simulator (HATS) to simulate the target speaker, and 8 surrounding loud speakers that simulate the sources of interference. Each interference speaker is spaced to have 45 degree gaps with set distance between 1\,m and 3\,m. For each speaker, two sets of speech data (ETSI American speech testset \cite{etsi_speech} and 8 speakers\footnote{p360, p361, p362, p363, p364, p374, p376, s5} in VCTK \cite{yamagishi_vctk_2019}) and two sets of noise data (ETSI noise \cite{etsi_noise} and DEMAND noise \cite{thiemann_deman_dataset_2013}) are played back and recorded, and the recordings are done separately for each speaker, and later mixed with different combinations and SNRs for beamformer derivation and model evaluation. The data collection process is repeated in three different room setups.

\vspace{-0.3cm}
\subsection{Beamforming}
\vspace{-0.2cm}
From the 3-microphone recordings, a causal linear time-invariant (LTI) STFT-domain beamformer is derived using a multi-frame multi-channel Wiener filter (MCWF) algorithm \cite{wang2021sequential}. Specifically, for each frequency bin, we obtain a linear combination of values from a causal window of 4 STFT frames across the 3 microphones. The covariance matrices required in the MCWF derivation are estimated using the ETSI speech and ETSI noise datasets recorded from two rooms synthesized at 3\,dB SNR. The beamforming coefficients derived from the covariance matrices of the first two rooms are then fixed, and the performance of the fixed beamformer on VCTK and DEMAND recorded in the third room\footnote{We make sure that the beamformer is tested on different datasets collected in a different room from the dataset used for its derivation so that the result is not overfitted to specific room acoustics or datasets.} is reported in Table~\ref{table:bss_sdr} (also in Fig.~\ref{fig:teaser}(b)). In Table~\ref{table:bss_sdr}, we mix the target speech recording from HATS with either interference speech or noise recordings from the 8 surrounding loudspeakers at 0\,dB and 6\,dB, and report the BSS-SDR score of the beamformer output compared against original speech played by HATS. Specifically the case of speech as interference at 0\,dB SNR\footnote{The target SNR is set taking all three microphones into account, and the SNR (or BSS-SDR) values can differ across the 3 individual microphones.}~is visualized in Fig.~\ref{fig:teaser}(b) in green as a polar plot. As can be observed from the data, there is about 3 to 4\,dB gain in BSS-SDR when the interference comes from 90\textdegree~and 270\textdegree ~angles, while the rejection gain is smaller at 0\textdegree~and 180\textdegree. This is the behavior expected from the beamformer, as it learns to preserve target speech at around 0\textdegree, and the three microphone are arranged in a plane which does not differentiate front and back.\footnote{Note that the BSS-SDR gain at 0\textdegree~is higher that of 180\textdegree, which is attributed to the fact the source of interference at 0\textdegree~is behind the HATS so the true angle of arrival of the interference signal deviates from 0\textdegree, allowing for slightly better rejection compared with 180\textdegree.} 

Note that our method would work with any fixed beamformer designed to keep the desired target direction and reject other directions, but we chose to estimate the beamformer from short recordings which match the real-world scenario, does not necessitate knowing the microphone geometry exactly, and implicitly takes care of device-related transfer functions which may be costly to estimate otherwise.

\subsection{Results and Discussion}
\label{subsec:result}

With the beamformer in place, we report the performance of GSENet and two baseline single-channel enhancement models SSENet(A) and SSENet(B) in Table~\ref{table:bss_sdr}. The experiment is set up such that: (1) GSENet takes both the beamformer output and one raw microphone waveform as inputs and compared with (2) SSENet is only exposed to the beamformer output as input. 

The left half of the table reports the BSS-SDR score when the target speech overlaps with an interfering speech signal from one of the 8 directions. At 0\,dB SNR of interfering speech, GSENet performs the best across the board, with as much as 8.5\,dB BSS-SDR gain on top of the beamformer output, and the gain correlates well with the beamformer performance. As is evident from the polar plot in Fig.~\ref{fig:teaser}(b), the larger the beamformer gain, the better contrast there is between the two input channels of GSENet, and with it, the easier it is for GSENet to identify and further suppress the interfering speech. In comparison, the single-channel models do not have enough context to differentiate the target speech from other directional interference and thus fail to provide satisfactory speech enhancement resulting in lower BSS-SDR scores. In the 6\,dB scenario, SSENet(B), which is trained to reject interfering speech with a lower signal strength when there is overlapping speech, delivers slightly better performance by 0.5\,dB compared with GSENet at 180\textdegree~ where there is not enough beamforming gain, but GSENet still outperforms SSENet(B) with an average of 2.4\,dB improvement for all the other directions. 

The right half of the table reports enhancement results with noise as interference. Since all three models are trained to recognize pattern of noise for denoising, both GSENet and its single channel counterparts perform similarly. In most of the cases though, GSENet still outperforms due to the contrast between the two input channels introduced by the beamformer.




\vspace{-0.4cm}
\section{Conclusion}
\vspace{-0.2cm}
\label{sec:conclusion}
In this work we proposed a multi-channel speech enhancement solution composed of a microphone-configuration specific beamformer and a device-agnostic speech enhancement model named GSENet that takes both the raw microphone audio and beamformer output as input. We proposed a model with low-latency real-time on-device inference capability and devised a training method that allows the model to learn from the contrast between the two input channels. Our solution delivers much better speech separation performance in the case of overlapping speech, as well as achieves similar or better denosing than baseline single-channel models, with as few as 3 microphones. In our future work we plan to extend the solution to benefit from the diversity provided by more than one beamformer to further improve the denoising and separation performance.

\vspace{-0.3cm}
\section{Acknowledgements}
\vspace{-0.2cm}
\label{sec:ack}
We would like to thank John Hershey, Artsiom Ablavatski, Yury Kartynnik, and Kevin Wilson for the feedback on this work. We also thank Jay Lin for the help on the recording setup.

\begin{table*}
\setlength{\tabcolsep}{3.6pt}
\caption{
 ViSQOL \cite{ViSQOL} score of the raw and enhanced speech waveform with interference coming from different angles. Left: speech as interference. Right: noise as interference. Top: interference at 0dB SNR. Bottom: interference at 6dB SNR.
}
\label{table:visqol}
\centering
\begin{tabular}{ @{}l c c c c c c c c c c c c c c c c c c c @{}} 
\toprule
& \multicolumn{9}{c}{Speech (VCTK) as interference} & \multicolumn{9}{c}{Noise (DEMAND) as interference}\\
 \cmidrule(lr){2-10} \cmidrule(lr){11-19}
Interference angle   & 0\textdegree & 45\textdegree & 90\textdegree & 135\textdegree & 180\textdegree & 225\textdegree & 270\textdegree & 315\textdegree & avg.  & 0\textdegree & 45\textdegree & 90\textdegree & 135\textdegree & 180\textdegree & 225\textdegree & 270\textdegree & 315\textdegree & avg.\\

\midrule
Reference Mic.  & 1.96 & 1.86 & 1.87 & 1.88 & 1.80 & 1.83 & 1.82 & 1.78 & 1.85 & 1.62 & 1.70 & 1.48 & 1.46 & 1.51 & 1.44 & 1.49 & 1.46 & 1.52\\ 
Beamformer(BFer) & 2.17 & 2.16 & 2.16 & 2.22 & 2.07 & 2.25 & 2.22 & 2.15 & 2.18 & 1.56 & 1.59 & 1.43 & 1.45 & 1.48 & 1.51 & 1.43 & 1.43 & 1.48\\ 
BFer + SSENet (A) & 2.22 & 2.21 & 2.21 & 2.26 & \se{2.11} & 2.29 & 2.28 & 2.20 & 2.22 & 2.04 & \se{2.18} & \se{2.02} & 2.00 & 1.77 & 2.07 & \se{2.05} & \se{1.96} & \se{2.01}\\
BFer + SSENet (B) & \se{2.41} & \se{2.44} & \se{2.39} & \se{2.46} & 1.99 & \se{2.45} & \se{2.47} & \se{2.44} & \se{2.38} & \se{2.06} & \be{2.19} & \be{2.05} & \se{2.02} & \be{1.82} & \se{2.09} & \be{2.08} & \be{1.99} & \be{2.04}\\ 
BFer + GSENet    & \be{2.79} & \be{3.01} & \be{3.02} & \be{3.01} & \be{2.27} & \be{3.00} & \be{3.05} & \be{3.01} & \be{2.90} & \be{2.07} & \be{2.19} & \se{2.02} & \be{2.03} & \se{1.78} & \be{2.15} & \be{2.08} & \be{1.99} & \be{2.04}\\ 

\midrule

Reference Mic.   & 2.42 & 2.34 & 2.30 & 2.36 & 2.23 & 2.32 & 2.28 & 2.25 & 2.31 & 1.91 & 2.00 & 1.73 & 1.74 & 1.73 & 1.77 & 1.77 & 1.75 & 1.80\\ 
Beamformer(BFer) & 2.53 & 2.55 & 2.54 & 2.54 & 2.38 & 2.56 & 2.56 & 2.53 & 2.53 & 1.76 & 1.79 & 1.64 & 1.58 & 1.67 & 1.66 & 1.69 & 1.63 & 1.68\\
BFer + SSENet (A) & 2.63 & 2.64 & 2.62 & 2.62 & 2.45 & 2.64 & 2.65 & 2.62 & 2.61 & \se{2.39} & 2.49 & 2.39 & 2.34 & 2.04 & 2.41 & 2.38 & 2.32 & 2.35\\ 
BFer + SSENet (B) & \se{3.00} & \se{3.02} & \se{2.99} & \se{3.03} & \be{2.71} & \se{3.01} & \se{3.02} & \se{2.99} & \se{2.97} & \be{2.42} & \se{2.52} & \se{2.42} & \se{2.36} & \be{2.10} & \se{2.43} & \se{2.42} & \se{2.35} & \se{2.38}\\
BFer + GSENet    & \be{3.06} & \be{3.27} & \be{3.26} & \be{3.24} & \se{2.63} & \be{3.24} & \be{3.30} & \be{3.27} & \be{3.16} & \be{2.42} & \be{2.55} & \be{2.43} & \be{2.42} & \se{2.07} & \be{2.51} & \be{2.47} & \be{2.41} & \be{2.41}\\

\bottomrule
\end{tabular}
\end{table*}

\let\oldbibliography\thebibliography
\renewcommand{\thebibliography}[1]{\oldbibliography{#1}
\setlength{\itemsep}{0pt}} 

\bibliographystyle{IEEEbib}
\bibliography{refs}

\appendix
\section*{Appendix}

\section{Additional evaluation results}

\begin{figure}[h]

     \includegraphics[width=0.5\textwidth]{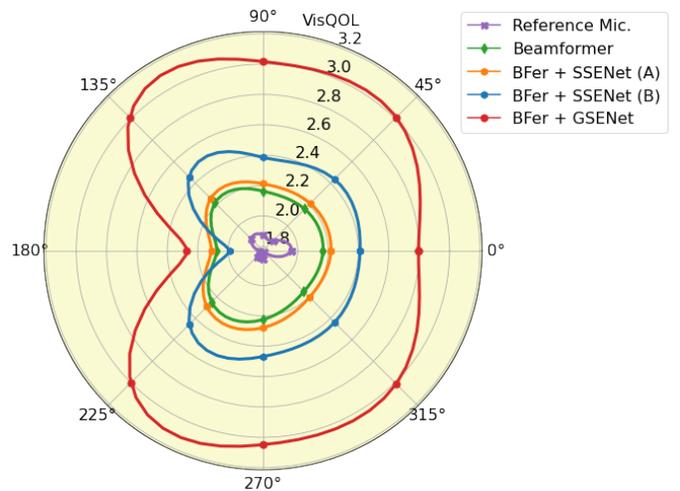}
     \caption{ViSQOL scores \cite{ViSQOL} in the case when interfering speech comes from 8 directions, each at 0\,dB SNR on a device with 3 microphones. Target speaker is at 0\,\textdegree. See Fig.~\ref{fig:listening_room} for the lab recording setup, Table~\ref{table:visqol} for more complete results.}
\vspace{-0.3cm}
\label{fig:visqol}
\end{figure}

In addition to the BSS-SDR evaluation results reported in Table~\ref{table:bss_sdr}, we also run evaluation using ViSQOL \cite{ViSQOL}, a score designed to be better aligned with perceptual quality, and report the results in Table~\ref{table:visqol}. The specific case of applying speech as interference at 0\,dB SNR is visualized in Fig.~\ref{fig:visqol} in a polar plot. Compared with BSS-SDR results (Table~\ref{table:bss_sdr} and Fig.~\ref{fig:teaser}(b)), similar observation can be made about the different solutions on ViSQOL (Table~\ref{table:visqol} and Fig.~\ref{fig:visqol}).

\end{document}